\documentclass[twocolumn,showpacs,floatfix,aps]{revtex4}
\usepackage{amssymb}
\usepackage[dvips]{graphics}

\def \beq {\begin{equation}}
\def \eeq {\end{equation}}

\begin{document}

\title{Non-minimal coupling, exponential potentials and the 
$w<-1$ regime of dark energy}

\author{F\'abio C. Carvalho}
\email{fabiocc@fma.if.usp.br}
\affiliation{
Instituto de F\'\i sica, Universidade de S\~ao Paulo,
CP 66318, 05315-970 S\~ao Paulo, SP, Brazil}

\author{Alberto Saa}
\email{asaa@ime.unicamp.br}
\affiliation{
IMECC, Universidade Estadual de Campinas,
C.P. 6065, 13083-859 Campinas, SP, Brazil.}


\begin{abstract}
Recent observations and theoretical considerations have motivated the
study of models for dark energy with equation of state characterized 
by a parameter $w=p/\rho<-1$. Such models, however, are usually believed
to be inviable due to their instabilities against classical perturbations
or potentially catastrophic vacuum decays. In this Brief Report, we
show that a simple quintessential model with potential 
$V(\phi)=Ae^{-\sigma\phi}$ and a gravitational
coupling of the form $(1+|\xi|\phi^2)R$ can exhibit, 
for large sets of initial conditions, 
 asymptotic de Sitter
behavior with $w<-1$ regimes. Nevertheless, 
the model is indeed stable at classical
and quantum level.
\end{abstract}

\maketitle

\section{Introduction}
The nature of the dark energy component responsible for the accelerated expansion
of the universe\cite{Riess,Perlmutter} 
is one of the most profound problem of Physics
(for a recent review, see \cite{Review}). 
The simplest way to describe dark matter
is by means of a cosmological constant $\Lambda$, which acts on the Einstein equations as an
isotropic and homogeneous source with equation of state
$-p_\Lambda=\rho_\Lambda=\Lambda$. Straightforward questions about possible fluctuations of dark energy leads
naturally to the introduction of a field $\phi(x)$ (the quintessence\cite{quintessence}) instead
of the cosmological constant $\Lambda$. Besides the issue of fluctuations, 
the field description is preferable since for some models (that ones, for instance,
provided with tracker solutions\cite{tracker}) the appearance of an accelerated
expansion (de Sitter) phase is a generic dynamical behavior, avoiding, consequently,
problems with fine-tunning of initial conditions in the early universe. Quintessential
equations of state are, in general, of the type $p_\phi=w\rho_\phi$, where the parameter $w$
can vary with time. For minimally coupled models with usual kinetic terms, one has
\beq
w = \frac{p_\phi}{\rho_\phi} = \frac{\dot{\phi}^2 - 2V(\phi)}{\dot{\phi}^2 + 2V(\phi)} \ge -1.
\eeq
According to the Einstein equations,
cosmological models with accelerated expansion phases 
require source terms for which $w<-1/3$. 

Recent observational questions\cite{obser} 
and theoretical speculations\cite{spec} have motivated the
analysis of the possibility of having realistic models for which, at least
temporally, $w<-1$. Indeed, recent data from Hubble Space Telescope of
Type Ia supernova at high redshifts
($z>1$)\cite{bigz} favorer scenarios with slowly evolving $w$ and restrict its values
to the range $w=-1.02\left(^{+0.13}_{-0.19}\right)$. 
They are compatible with a simple cosmological constant $(w=-1)$, but if, however,
 one really has $w<-1$, both
cosmological constant and minimally coupled scalar quintessence
descriptions for dark energy are ruled out. Realistic models accommodating also the
regime $w<-1$ are, of course, welcome.

In\cite{CHT}, Carroll, Hoffman and Trodden consider the viability of constructing
realistic dark energy models with $w<-1$. The so called phantom fields, {\em i.e.}
minimally coupled scalar 
fields with ``negative" kinetic energy, are usually invoked to construct such
models. Despite that the dominant energy condition (in fact, {\em all} usual energy
conditions) is violated in models with $w<-1$, the authors are able to introduce a classically
stable model involving a phantom field. However, due to the peculiar kinetic energy
of the phantom, the model is unstable against any quantum process involving it. In particular,
as the phantom 
potential is unbounded from below, there are catastrophic vacuum decays\cite{CHT}.
Furthermore, 
some general results\cite{grad} suggest that any minimally coupled theory with $w<-1$ has
spatial gradient instabilities that would be ruled out by CMB observations.
These results put severe doubts on the viability of dark energy models based on
phantom fields.

In this Brief Report, we notice that certain quintessential
models can exhibit a generic asymptotic de Sitter phase with many solutions
for which $w$ is slowly evolving and lesser than -1, without the introduction of any 
(classical or quantum)
instability. The quintessential field is assumed to be non-minimally coupled to gravity
\beq
\label{act}
S=\int d^4x \sqrt{-g}\left\{F(\phi)R - \partial_a\phi\partial^a\phi
-2V(\phi) \right\},
\eeq
where $F(\phi)=1-\xi\phi^2$, 
  $\xi<0$, 
and to have an exponential self-interaction potential
\beq
\label{pot}
V(\phi) = Ae^{-\sigma\phi}.
\eeq
Exponential potentials have been used recently in cosmology, mainly in
connection with tracker fields\cite{expon}.
They appear naturally in higher-dimensional theories and string inspired models\cite{higher}.
Non-minimally coupled models are also commonly adopted.  
In \cite{PRD}, for instance, a quintessential
model with $\xi=1/6$ (conformal coupling) and 
$V(\phi)=\frac{m}{2}\phi^2-\frac{\Omega}{4}\phi^4$
was introduced. With such potential, however, 
conformally coupled models never exhibit asymptotic
de Sitter phases. Moreover,   models with coupling $F(\phi)R$ are {\em generically}
singular on the hypersurfaces $F(\phi)=0$\cite{PRD2}, precluding the viability of
quintessential models with conformal coupling. Some proposals to circumvent this
singularity with the inclusion of higher order gravitational terms in the
action has been suggested\cite{GS}, but the viability of the resulting models is still
unclear. In this work, we avoid this  singularity by choosing hereafter $\xi<0$.
Non-minimal couplings of the type $F(\phi)=1+|\xi|(\phi^2-\phi_0^2)$
has been   considered in \cite{mata} to study tracking behavior
for potentials $V(\phi)=M^{4+\alpha}/\phi^\alpha$, $\alpha>0$. The case of
$F(\phi)=|\xi|\phi^2$ was considered in \cite{bartolo}.
As we will see, despite that the model given by (\ref{act}) is known to violate
weak energy condition\cite{FT}, the phase space for the comoving frame presents large
regions of stability.

\section{The model}

The Einstein equations obtained from the action (\ref{act}) are
\beq
\label{e1}
FG_{ab}= \nabla_a\phi\nabla_b\phi - \frac{g_{ab}}{2} \left( 
\nabla_c\phi\nabla^c\phi + 2V - 2\Box F
\right)+ \nabla_a\nabla_b F, 
\eeq
while the Klein-Gordon equation is
\beq
\label{e2}
\Box\phi -V'+ \frac{1}{2}F'R=0.
\eeq
A relevant issue here is how to define an equation of state for the field
$\phi$ from the equations (\ref{e1}) and (\ref{e2}). The r.h.s
of (\ref{e1}) does not correspond to a covariantly conserved energy momentum
tensor due to the presence of $F(\phi)$ in its l.h.s. In order to define
the pressure and energy for the field $\phi$ in a consistent way with the
continuity equation, one needs a covariantly conserved energy momentum tensor,
and from (\ref{e1}) we can get $G_{ab} = T_{ab}$ with\cite{Torres}
\beq
\label{tab}
T_{ab} = \nabla_a\phi\nabla_b\phi - \frac{g_{ab}}{2} \left( 
\nabla_c\phi\nabla^c\phi + 2V - 2\Box F
\right) + \nabla_a\nabla_b F + (1-F)G_{ab}.
\eeq

Assuming an isotropic and spatially flat universe,
\beq
\label{metric}
ds^2 = -dt^2 + a^2(t)\left( dx^2 + dy^2 + dz^2 \right),
\eeq
we get, from the temporal component of the Einstein equation (\ref{e1}), the energy constraint  
\beq 
\label{energy}
3H\left(FH + F'\dot\phi\right) = \frac{\dot{\phi}^2}{2} + V(\phi),
\eeq
where $H=\dot{a}/a$.
From the spatial components, one gets
the modified Friedmann equation
\beq
\label{einstein}
-2F_1\dot H = 3 \left( F + 2(F')^2 \right)H^2 +
\frac{1+2F''}{2}{\dot\phi}^2 
 - V - \left(H\dot\phi +V' \right)F',
\eeq
where $F_1(\phi) = F(\phi) + \frac{3}{2}\left(F'(\phi) \right)^2$.
For the metric (\ref{metric}), the Klein Gordon equation reads
\beq
\label{KG}
\ddot\phi + \frac{G(\phi,\dot\phi,H)}{F_1}\dot\phi + V'_{\rm eff}(\phi)=0,
\eeq
where
\beq
\label{G}
G(\phi,\dot\phi,H) = 3F_1H + \frac{1}{2}(1+3F'') F' \dot\phi 
\eeq
and
\beq
\label{veff}
V'_{\rm eff}(\phi) = \frac{1}{F_1} \left(FV'-2F'V \right) 
\eeq
The fixed points of the equations (\ref{energy})-(\ref{KG}) are the  constant solutions
$\phi(t)=\bar\phi$ and $H(t)=\bar H$,  corresponding to de Sitter solutions 
$a(t)\propto e^{\bar Ht}$ for which $w=-1$.
Despite that the potential (\ref{pot}) has no equilibrium
points, thanks to the non-minimal coupling, the model has indeed the fixed points 
$(\bar\phi_-,\pm \bar H_-)$ and 
$(\bar\phi_+,\pm \bar H_+)$, where
\beq
\bar\phi_\pm = -\frac{2}{\sigma}\left(1\mp\sqrt{1 + \sigma^2/4\xi} \right)
\eeq
and
\beq
\bar H_\pm^2 = \frac{V(\bar\phi_\pm)}{3F\left(\bar\phi_\pm\right)}.
\eeq
The fixed points exist, of course, only for $\xi\le -\sigma^2/4$.
The next Section is devoted to the study of the phase space of this model.
As we will see, the fixed point $(\bar\phi_-, \bar H_-)$ is an attractor;
large sets of 
solutions tend spontaneously to this de Sitter phase, irrespective
of their initial conditions. When approaching the de Sitter point,
  solutions can have $w<-1$, without any induced instability. 

We finish this Section with the definition of the pressure $p_\phi$ and
energy $\rho_\phi$. These quantities are defined from the energy momentum
tensor as $T_{ab} = (\rho+p)u_au_b + pg_{ab}$,
where $u_a $ is a globally timelike vector.
With the hypothesis of isotropy and homogeneity, we
get, in the comoving frame,
$\rho_\phi = 3H^2$ and $p_\phi = -\left(2\dot H +3H^2 \right)$, leading to
\beq
\label{w}
w = - \left( 1+\frac{2}{3}\frac{\dot H}{H^2} \right).
\eeq
As one can see, the ratio $w$ defined in (\ref{w}) is not  subject to the 
restriction  $w\ge-1$. Note that the continuity equation
$\dot\rho_\phi + 3H (\rho_\phi + p_\phi) = 0$,
which is a direct consequence of the Bianchi identities
for the Einstein equations, holds here.

\section{The phase space}

\begin{figure}[ht]
\resizebox{\linewidth}{!}{\includegraphics*{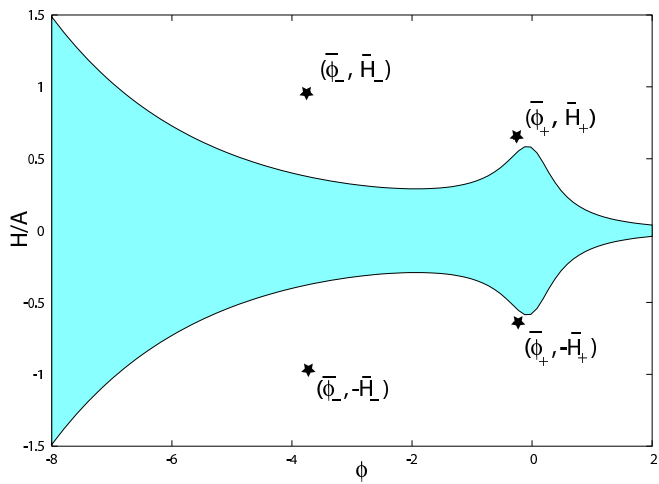}}
\caption{The fixed points  for the equations
(\ref{energy})-(\ref{KG}) on the plane 
$(\phi,H)$. The shadowed region ($H^2<V/3F_1$) is dynamically unaccessible.
For this graphics,
$-\xi=\sigma=1$.}
\label{fig1}
\end{figure}

We study the phase space of the model by using the same semi-analytical approach used
in \cite{PRD}. The phase space is 3-dimensional $(\phi,\dot\phi,H)$, but due to
the energy constraint (\ref{energy}) the dynamics are restricted to a 2-dimensional
submanifold. No restrictions are imposed
on the $(\phi,\dot{\phi})$ phase portrait,
but for the $(\phi,H)$ one, only the region 
$H^2\ge V/3F_1$ is dynamically allowed (See Fig. \ref{fig1}).
Since $H=0$ is not allowed by the dynamics, the trajectories
are confined to semi-spaces $H>0$ and $H<0$. We are concerned here only with 
$H>0$. 
We remind that real trajectories move
on the 2-dimensional manifold defined by the energy constraint (\ref{energy}).
Therefore, in the projection on the plane 
$(\phi,H)$, each point on the allowed region corresponds, in fact, to two  
possible values for $\dot\phi$ (two ``sheets"), with the exception of the lines $H^2 = V/3F_1$,
where only one value for $\dot\phi$ is allowed. Solutions aways ``cross" from one
sheet to another tangentially to the lines $H^2 = V/3F_1$.

The aspect of $V_{\rm eff}(\phi)$ obtained from (\ref{veff}) is crucial to the
identification of the attractor points in the phase space. From (\ref{veff}), one has
\beq
V'_{\rm eff}(\phi) = A\sigma\xi e^{-\sigma\phi}
\frac{\left(\phi-\bar\phi_+ \right)\left(\phi-\bar\phi_- \right)}{1-\xi(1-6\xi)\phi^2},
\eeq
which can be integrated by parts in terms of the Exponential Integral function
$Ei(x)$\cite{abram}.
\begin{figure}[ht]
\resizebox{\linewidth}{!}{\includegraphics*{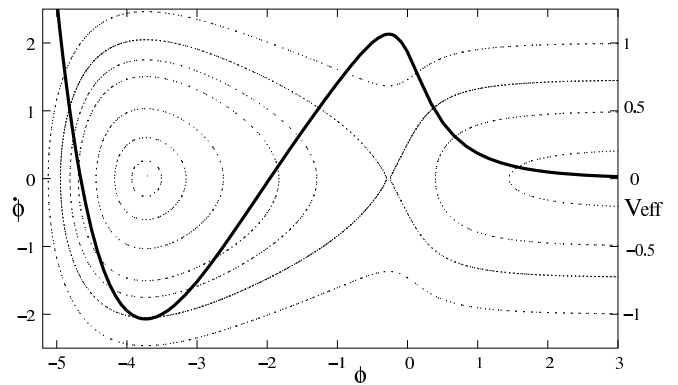}}
\caption{Aspect of the potential (the solid line) $V_{\rm eff}(\phi)$ for $-\xi=\sigma=1$.
The contours correspond to the lines of constant $L(\phi,\dot\phi)$ 
(\ref{lyap}). Since they
are closed around the fixed point $\bar\phi_-$, the condition $\dot L < 0$
along solutions implies that they tend asymptotically to $\bar\phi_-$.
}
\label{fig2}
\end{figure}
From Fig. \ref{fig2}, it is clear that the fixed point corresponding to
$\bar\phi_+$ is unstable, while $\bar\phi_-$ should correspond to a stable one.
However, despite the clear fact that $\bar\phi_-$ is a
minimum of $V_{\rm eff}$, one cannot conclude safely that it indeed corresponds to
a attractor due to the function $G(\phi,\dot\phi, H)$ given by (\ref{G}). If
$G(\phi,\dot\phi, H)\ge 0$, the solutions of the Klein-Gordon equation  
around the point $\bar\phi_-$ will be simple dumped oscillations. This can
be checked by introducing the function 
\beq
\label{lyap}
L(\phi,\dot\phi) = \frac{\dot\phi^2}{2} + V_{\rm eff}(\phi),
\eeq
and noticing that $\dot L = - G\dot\phi^2$ along the solutions of the Klein-Gordon
equation (\ref{KG}). Provided that $G(\phi,\dot\phi, H)\ge 0$, the function
$L(\phi,\dot\phi) $ is a Lyapunov function\cite{PRD} for the fixed point 
$\bar\phi_-$, assuring its stability. 
From the Eq. (\ref{einstein}), we see that when $\phi(t)$ approaches $\bar\phi_-$,
$\dot H\rightarrow 0$ and $H(t)$ approaches $\bar H_-$, establishing the attractor
character of the fixed point $(\bar\phi_-, \bar H_-)$.

From (\ref{G}), we have
\beq
\label{G1}
G(\phi,\dot\phi, H) = 3 \left( 
F(\phi)H + \frac{1-6\xi}{6}F'(\phi)\dot\phi
\right) + \frac{9}{2} \left( F'(\phi) \right)^2 H,
\eeq
and from the energy constraint (\ref{energy}), one has that $F(\phi)H+F'(\phi)\dot\phi>0$ 
on the semi-space $H>0$, implying the positivity of $G$  and, hence,
establishing the attractive character of the fixed point $(\bar\phi_-,\bar H_-)$, for $\xi > -5/6$. 
However,
this is a very conservative lower bound for $\xi$. Our exhaustive numerical simulations
\begin{figure}[ht]
\resizebox{\linewidth}{!}{\includegraphics*{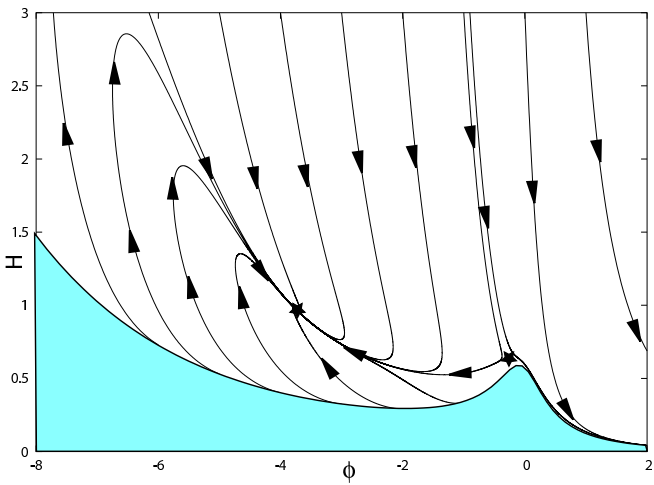}}
\caption{Typical phase portrait $(\phi, H)$, corresponding to the case
$-\xi=A=\sigma=1$.  Due to the symmetries of
Eq. (\ref{energy})-(\ref{KG}), the $H<0$ trajectories are obtained 
from the $H>0$ ones by a reflexion on $H=0$ and reversion of the arrows (time reversal operation). }
\label{fig4}
\end{figure}
suggest that it can be considerably smaller. We could  verify the attractive
character of the fixed point $(\bar\phi_-, \bar H_-)$ even for $\xi<-100$,
suggesting that eventual amplifications forthcoming from the  $G(\phi,\dot\phi,H)<0$
regions are not enough to win the potential $V_{\rm eff}$ around $\bar\phi_-$.
 A typical
phase portrait is displayed in Fig. \ref{fig4}, corresponding to the case
$-\xi=A=\sigma=1$. The attraction basin is considerably
larger than the conservative estimative base on the closed lines of constant $L$ around
the fixed point. Note that all solutions starting with $\phi>\bar\phi_+$ are runaway solutions.
\begin{figure}[ht]
\resizebox{\linewidth}{!}{\includegraphics*{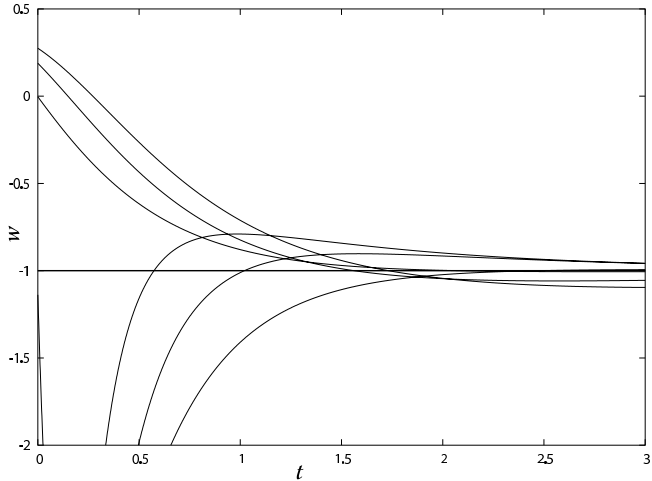}}
\caption{The parameter $w$ along some solutions presented in
the phase portrait of Fig. \ref{fig4}.
According to (\ref{w}), solutions have $w< -1$ when $\dot{H}> 0$,
and $w> -1$ when $\dot{H}< 0$.}
\label{fig5}
\end{figure}
Fig. \ref{fig5} shows the curves $w(t)$ for some solutions presented in Fig. \ref{fig4}.
According to (\ref{w}), solutions approaching the fixed point from below have $w<-1$, while that ones doing
from above have  $w>-1$.

\section{Conclusion}

We show here that non-minimally coupled quintessential models with exponential
potentials can exhibit asymptotic de Sitter phases for large sets of initial 
conditions. Some of these phases are characterized by a slowly evolving parameter 
$w=p_\phi/\rho_\phi <-1$, compatible, in principle, with the recent observational
data. Our 
analysis is based on the existence of a Lyapunov function for the Klein-Gordon
equation that can be used to estimate the attraction basin of the relevant
fixed points. As it was already mentioned, real attraction basins are typically
much larger than these estimations.
Our results are in agreement with the linearized analysis recently
proposed in \cite{faraoni}.

We stress that the model presented here is free from the instabilities that are
usually associated to phantom models. Classically, since $F(\phi)\ne 0$ , the
model is not plagued with the anisotropic singularities described in \cite{PRD2}.
Besides, since $F(\phi)$ is always positive and $V(\phi)$ is bounded from below,
the model is also free from the quantum instabilities described in \cite{CHT}
for phantom fields.

We conclude, therefore, that it is possible, in principle, to construct
realistic models for dark energy with $w<-1$. Relevant
issues now are the introduction of matter fields and the study of the role
played by the strong non-minimal coupling regime\cite{bartolo,strong} in the model.
These points are under investigation.

\acknowledgments

This work was supported by FAPESP and CAPES.

\end{document}